\documentclass[10pt]{elsart}
\usepackage{graphicx}
\begin{document}
\begin{frontmatter}
\title{Coherent transport in linear arrays of quantum dots: the effects of
period doubling and of quasi-periodicity}
\author{M. R. Bakhtiari},
\author{P. Vignolo\corauthref{cor1}}, 
\author{M.P. Tosi}
\corauth[cor1]{Corresponding author, e-mail: {\tt vignolo@sns.it}}
\address{NEST-INFM and Classe di Scienze, Scuola Normale
Superiore, I-56126 Pisa, Italy}
\maketitle

%$\vspace{7cm}$

\begin{abstract}
We evaluate the phase-coherent transport of electrons along linear structures
of varying length, which are made from two types of potential wells set in
either a periodic or a Fibonacci quasi-periodic sequence. The array is
described by a tight-binding Hamiltonian and is reduced to an effective 
dimer by means of a decimation-renormalization method,
extended to allow for connection to external metallic leads, and the 
transmission coefficient is evaluated in a $T$-matrix scattering approach.
Parallel behaviors are found for the energy dependence of the density of 
electron states and of the transmittivity of the array. In particular,
we explicitly show that on increasing its length the periodic
array undergoes a metal-insulator transition near single occupancy per dot,
whereas prominent pseudo-gaps emerge away from the band center in the 
Fibonacci-ordered array.
\end{abstract}

\end{frontmatter}

\newpage
\section{Introduction}
The study of electron transport in mesoscopic systems is 
a fundamental problem in nanophysics and nanotechnology \cite{datta,ahmed}
and rests on Landauer's idea that conductance
is proportional to phase-coherent electron transmittivity \cite{landauer}.
Within this context a great deal of attention has been given in recent
years to electronic  transport in a variety of structures including
polymeric and molecular wires, nanotubes and quantum wires, quantum dots,
and arrays and networks of quantum dots.

A semiconductor quantum dot (QD) is often described as an 
artificial atom, in which an external potential replaces the 
attractive potential of the nucleus to confine
charge carriers in all three 
spatial directions \cite{jacak,ref2}.
The dot size is of the order of the Fermi wavelength 
in the host material,
typically between 10 nm and 1$\mu$m, and 
the confinement is usually achieved by 
electrical gating of a two-dimensional electron gas, often in 
combinations with etching of the material. 
Precise control of the number 
of electrons in the conduction band of a QD has
been achieved in GaAs heterostructures \cite{tarucha}. 
An all-electrically controlled 
quantum dot array (QDA) can be realized
by electrodes confining single
electrons to the dot regions, with the electron tunnel between neighboring 
dots being controlled by electrical gating \cite{Burkard}.

As in the case of real clusters,
electronic transport through QDA's can be treated theoretically
within a tight-binding framework \cite{tews}. 
The tight-binding model 
has been used to study 
Coulomb effects on QD transmittance \cite{aldea},
magnetoconductance in chaotic arrays \cite{louis},
Kondo resonances and Fano antiresonances in transport 
\cite{torio},
and transport through two-dimensional networks
\cite{dorn,Kirczenow,tanatar}.
A simplifying assumption treats a QD as a potential well, thus
omitting a detailed account of its internal structure.
This allows analytical results to be obtained for transport
properties of QDA's (see for instance \cite{mardani,mardani2}). 

In this work we use the simple tight-binding approach to derive
an analytical treatment of phase-coherent transport through a QDA model
composed of two different sets of QD's, as originally proposed by
Mardaani and Esfarjani \cite{mardani}. We treat both a periodic and 
quasi-periodic arrangement of the two types of QD's in the array,
and report comparative numerical illustrations of
the density-of-states (DOS) 
and of the transmittivity of these arrays for a number $N$ of QD's up
to 100. Our analytical results allow us to evaluate very long arrays
without any special numerical effort.
Specifically, we 
use a decimation-renormalization method~\cite{grosso,patty}
to reduce the QDA to an
effective dimer with renormalized site energies
and hopping interactions, and supplement the method with a
generalized Kirkman-Pendry relation \cite{kirkman,patty_k} 
to account for the connection of the QDA to external metallic leads.  
The $T$-matrix formalism introduced in this paper 
for the calculation of the QDA transmittivity
is equivalent to the formalism given in Ref. \cite{zehra}
for out-of-equilibrium leads in
the special case of an infinitesimal bias and 
is suitable to 
deal with renormalized effective Hamiltonians 
describing very long chains.
\section{Hamiltonian and density-of-states}
\label{H_e_dos}
We use a one-dimensional (1D) tight-binding Hamiltonian to describe a QDA of
$N$ equally spaced potential wells. The Hamiltonian can be written as
\begin{equation}\label{eq:H}
H_{\textrm{QD}}=\sum_{i=1}^N\hspace{2mm}\{e_i\,
|\,i\rangle\,\langle i\,|\,+(t_{i,i+1}\,|\,i\rangle\,\langle
i+1|+t_{i+1,i}|\,i+1\rangle\,\langle i\,|\,) \, \}\,,
\end{equation}
where the site energy $e_i$ corresponds to the energy level 
for an electron in the $i$-th well
and $t_{i,i+1}=t_{i+1,i}$ is the hopping energy between the 
$i$-th and the $(i+1)$-th well. The connection to external leds will 
be added later.

\begin{figure}
\begin{center}
\includegraphics[width=0.9\linewidth]{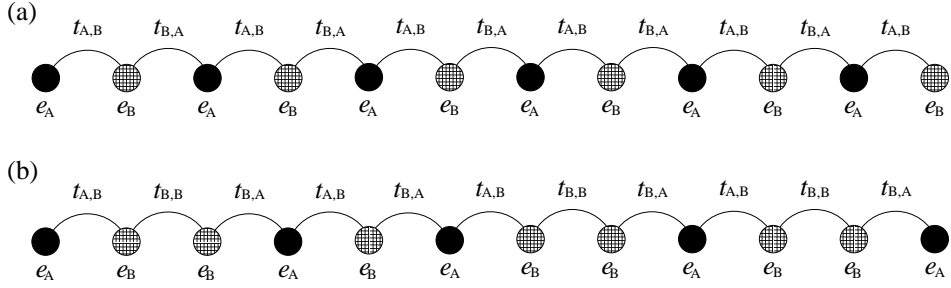}
\caption{Schematic Hamiltonian for (a) a periodic QDA 
and (b) a Fibonacci QDA, for the case $N=12$. \label{fig_schem}}
\end{center}
\end{figure}

We focus on two types of QDA's.
The first is a periodic array in which two
different types of QD's, $A$ and $B$ say, alternate.
In the second the two
kinds of QD's are arranged according to the Fibonacci
sequence, i.e., if we define the first QD as $F_1=A$ and the
second as $F_2=B$, the rest of the chain is built 
by the rule $F_n=F_{n-1}\, F_{n-2}$. 
The schematic Hamiltonians for these QDA's
are shown in Fig. \ref{fig_schem}.
  
The related Green's function
at energy $E$ is
\begin{equation}
G(E)=\frac{1}{E-H_{\textrm{QD}}}.
\end{equation}
Here and below, we consider a complex energy variable defined
as $E+i\,\eta$ in the
limit of vanishing positive imaginary part. 
To evaluate the DOS $D(E)$ we use
the Kirkman-Pendry relation~\cite{kirkman}, which relates 
the DOS to the
matrix element $G_{1,N}(E)$ expressing
coherence  between the first and the last site of the QDA,
\begin{equation}\label{eq:DOS}
D(E)=\frac{1}{\pi}\,\, {\textrm{Im}} \, \frac{\partial }{\partial
E}\ln G_{1,N}(E)\,.
\end{equation}
The matrix element $G_{1,N}(E)$ can be evaluated by reducing the 
QDA to an effective dimer through decimation of the intermediate QD's
(see for instance \cite{grosso,patty}).
The renormalized array contains just the first and the last
site and its Hamiltonian is expressed as a  $2\times 2$
matrix,
\begin{equation}
\tilde{H}_{\textrm{QD}}(E)=
 \left(
\begin{array}{ccc}
\tilde{\varepsilon}_1^{(N-2)}(E) & & \tilde{t}_{1,N}(E) \\
 \tilde{t}_{N,1}(E)& &\tilde{\varepsilon}_N^{(N-2)}(E)\\
\end{array}
\right).
\end{equation}
Here, the effective site and hopping energies can be
obtained by the recursive relations
\begin{eqnarray}
\tilde{\varepsilon}_1^{\,(j)}(E)&=&\tilde{\varepsilon}_1^{\,(j-1)}(E)+
\tilde{t}_{1,j+1}(E)\,\frac{1}{E-\tilde{\varepsilon}_{j+1}^{\,(j-1)}(E)}
\,\,\,t_{j+1,j+2}\,,\label{eq:rec1}\\
\tilde{\varepsilon}_{j+2}^{\,(j)}(E)&=&e_{j+2}+t_{j+2,j+1}
\,\frac{1}{E-\tilde{\varepsilon}_{j+1}^{\,(j-1)}(E)}\,\,
\tilde t_{j+1,1}(E)\,,\\
\tilde{t}_{1,j+2}(E)&=&\tilde{t}_{1,j+1}(E)\,
\frac{1}{E-\tilde{\varepsilon}_{j+1}^{\,(j-1)}(E)}\,\,\,t_{j+1,j+2}
\label{eq:rec3}
\end{eqnarray}
and $\tilde t_{j+1,1}=\tilde t_{1,j+1}$ for $j\ge 1$, the initial values
being given by the Hamiltonian parameters namely 
$\tilde{\varepsilon}_i^{\,(0)}(E)=e_i$ and $\tilde t_{1,2}(E)=t_{1,2}$.
By direct inversion of $(E-\tilde{H}_{\textrm{QD}}(E))$ we obtain
\begin{equation}
G_{1,N}(E)=\frac{\tilde{t}_{N,1}(E)}
{[E-\tilde{\varepsilon}_1^{(N-2)}(E)]
[E-\tilde{\varepsilon}_N^{(N-2)}(E)]-[\tilde{t}_{1,N}(E)]^2}.
\end{equation}
We then use Eq.~(\ref{eq:DOS}) to calculate the DOS.

\begin{figure}
\begin{center}
\includegraphics[width=0.49\linewidth]{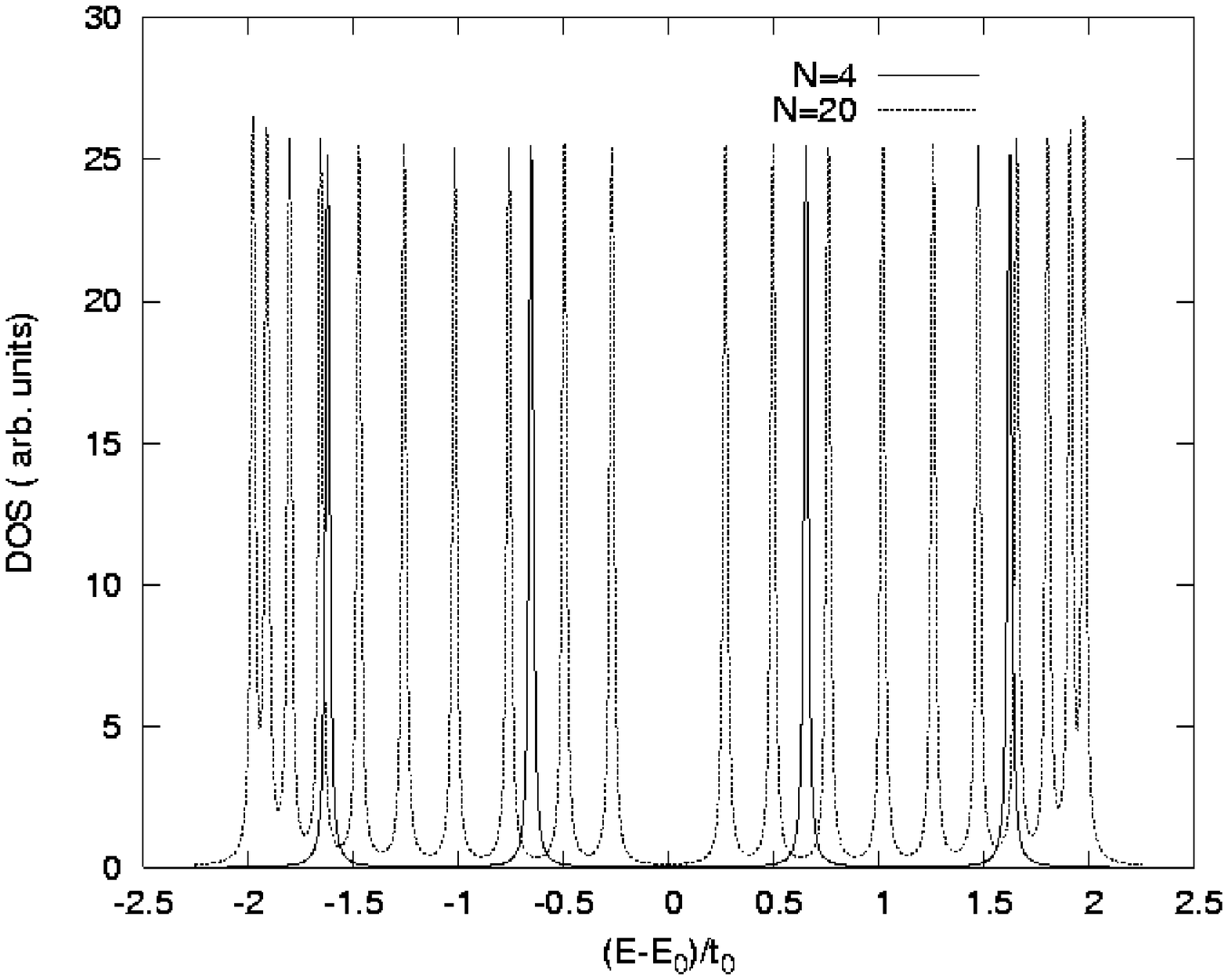}
\includegraphics[width=0.49\linewidth]{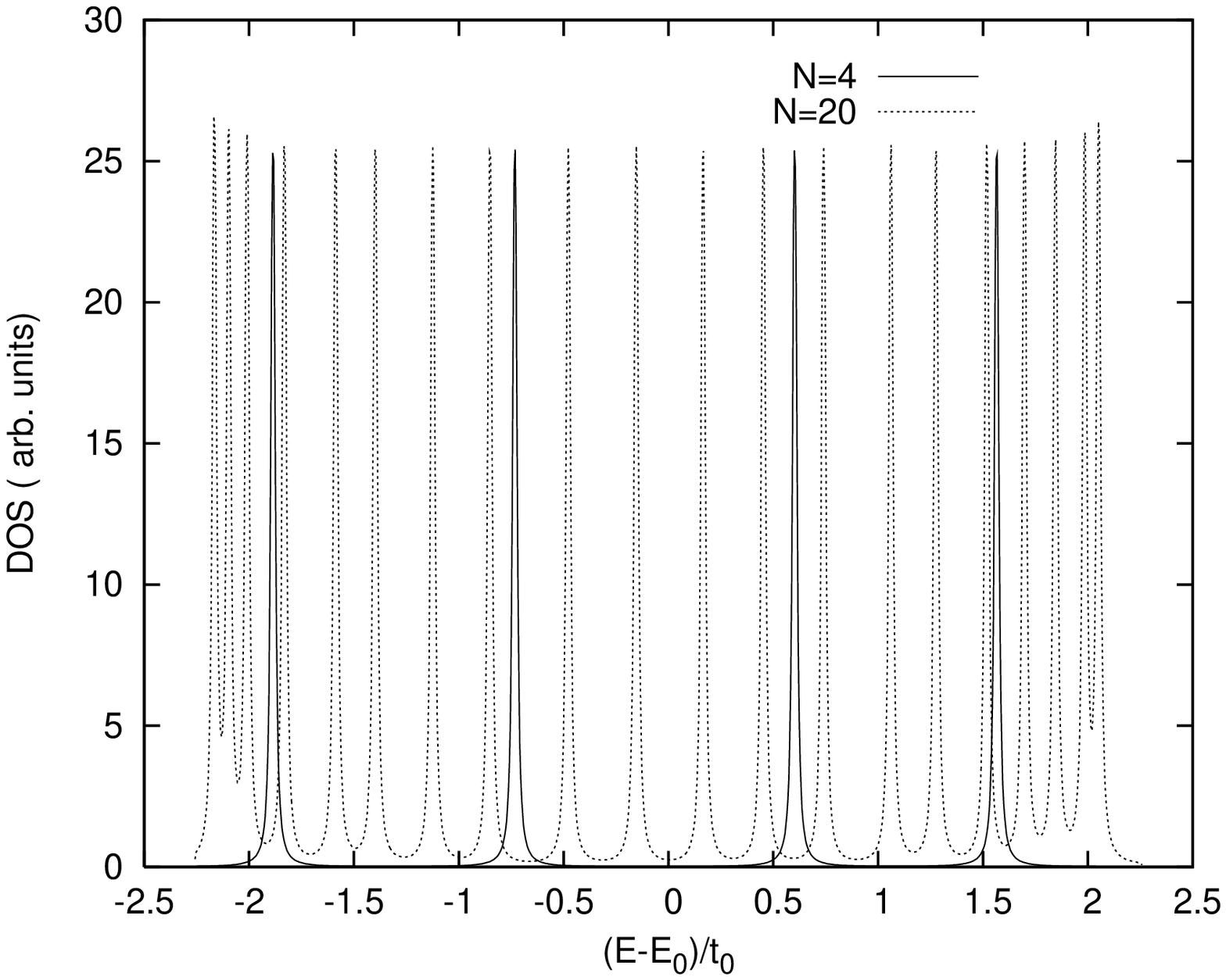}
\caption{DOS of periodic array (left) and Fibonacci-ordered
array (right) for the case $N=4$ (solid line) and $N=20$ (dotted line),
as a function of energy
$E$ referred to the center of the spectrum at energy $E_0$ and for a 
spectral width of $4t_0$.\label{fig:dos1}}
\end{center}
\end{figure}

\begin{figure}
\begin{center}
\includegraphics[width=0.49\linewidth]{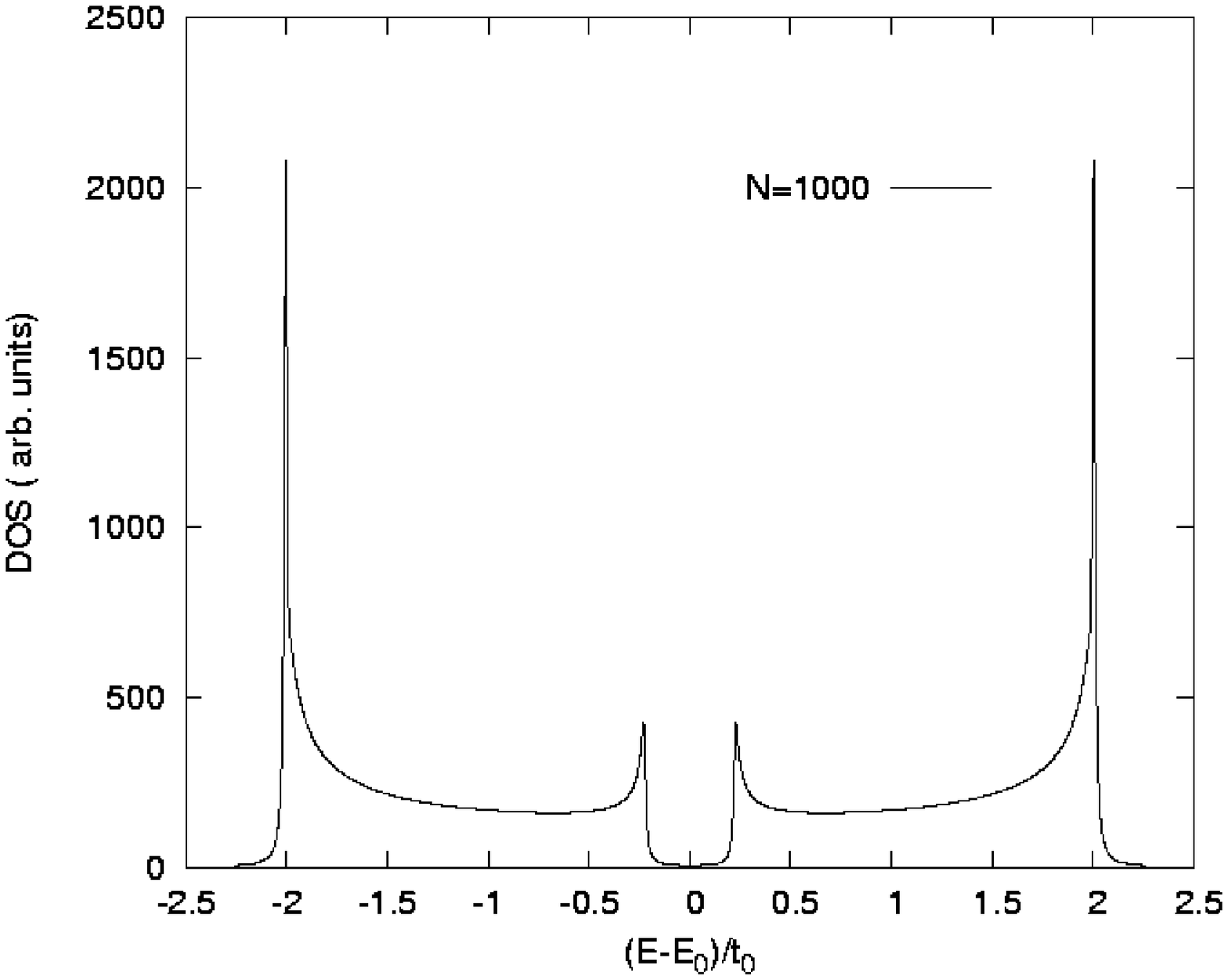}
\includegraphics[width=0.49\linewidth]{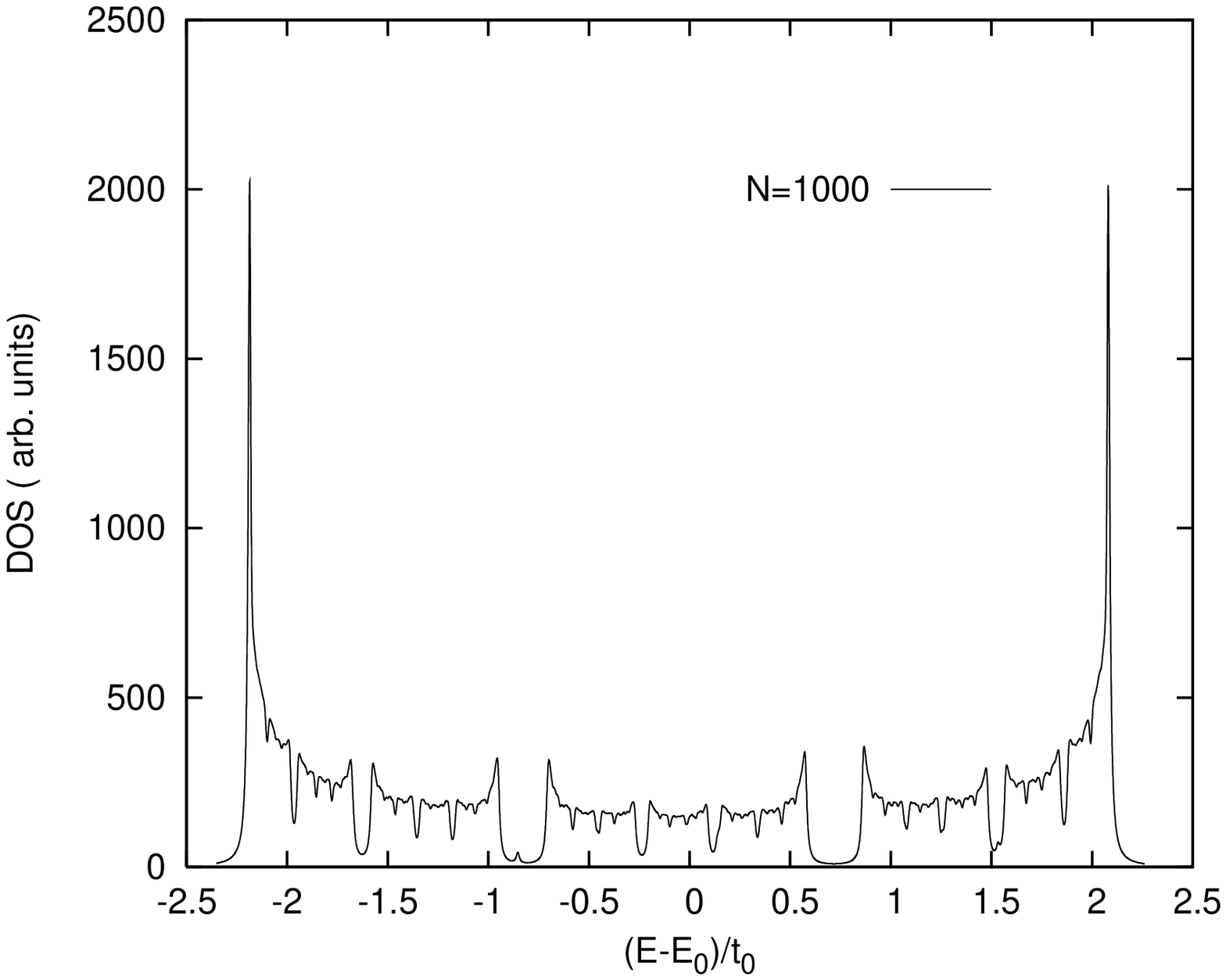}
\caption{DOS as in Fig. \ref{fig:dos1}, for the case $N=1000$.
\label{fig:dos2}}
\end{center}
\end{figure}

In the illustrative
calculations that we report here and in the following sections
we have used
the values of the parameters  
$e_A=-0.25$ eV, $e_B=0.25$ eV, $t_{A,B}=1.1$ eV and $t_{B,B}=1.0$ eV.
In Fig. \ref{fig:dos1} and
Fig. \ref{fig:dos2} we show the DOS for QDA's made from various
numbers of QD's in a periodic and a Fibonacci
configuration.
For small values of $N$ (see Fig. \ref{fig:dos1})
the DOS shows $N$ peaks distributed over the energy range.
For long chains with $N=1000$ (see Fig. \ref{fig:dos2})
the DOS is similar to that of an infinite array:
in the periodic sequence (left panel in Fig. \ref{fig:dos2})
there are two sub-bands separated by an energy gap of width $|e_A-e_B|$,
while in
the case of a Fibonacci array (right panel in Fig. \ref{fig:dos2})
the quasi-periodicity
induces a fragmentation of the spectrum and the appearance of pseudo-gaps.

\subsection{The effects of the leads}
The QDA is next connected to
an incoming ($l$) and an
outgoing ($r$) metallic lead, which are described by two
additional terms in the Hamiltonian,
\begin{equation}
H_{\textrm{L,l}}=\sum_{n=-\infty}^1\,\{E_0|n\rangle\langle n|+
t_0(|n-1\rangle\langle n|+\textrm{c.c.})\}
\end{equation}
and
\begin{equation}
H_{\textrm{L,r}}=\sum_{n=N}^\infty\,\{E_0|n\rangle\langle n|+
t_0(|n\rangle\langle n+1|+\textrm{c.c.})\}\,.
\end{equation}
Here the site energy $E_0$ in the leads and the hopping energy $t_0$ have been
chosen equal to the center of the spectrum of the QDA and to one
fourth of its spectral width.
The presence of the leads modifies the DOS of the system and
thus affects the electronic transport through the QDA.
We accordingly have to
decimate the leads and to consistently renormalize the energies of sites 1 and 
$N$ in the QDA. 

Equation (\ref{eq:DOS}) does not apply 
in general and in particular, when 
sites 1 and $N$ in the array
are not edge sites,
a modified 
Kirkman-Pendry relation as derived by 
Farchioni {\it et al.} \cite{patty_k} must instead be used. 
This relation is
\begin{equation}
D(E)=\frac{1}{\pi}\,\, {\textrm{Im}} \, \left.\frac{\partial }{\partial
\lambda}\ln G_{1,N}(E,\lambda)\right|_{\lambda=0},
\end{equation}
where the real parameter $\lambda$ is introduced to select the sites
on which the electronic states are counted.
In our specific case $G_{1,N}(E,\lambda)$ can be written as
\begin{equation}
\hspace{-0.7cm}
G_{1,N}(E,\lambda)=\frac{\tilde{t}_{N,1}(\tilde E)}
{[\tilde E-\tilde{\varepsilon}_1^{(N\!-2)}
(\tilde E)-\tilde{\mathcal E}(E)]
[\tilde E-\tilde{\varepsilon}_N^{(N\!-2)}
(\tilde E)-\tilde{\mathcal E}(E)]-
[\tilde{t}_{1,N}(\tilde E)]^2},
\end{equation}
where $\tilde E=E+\lambda$ and
the term $\tilde{\mathcal E}(E)=\frac{1}{2}(E-E_0)-
[\frac{1}{4}(E-E_0)^2-t_0^2]^{1/2}$
is the renormalized contribution of the leads
to the energy of sites 1 and $N$.

\begin{figure}
\begin{center}
\includegraphics[width=0.49\linewidth]{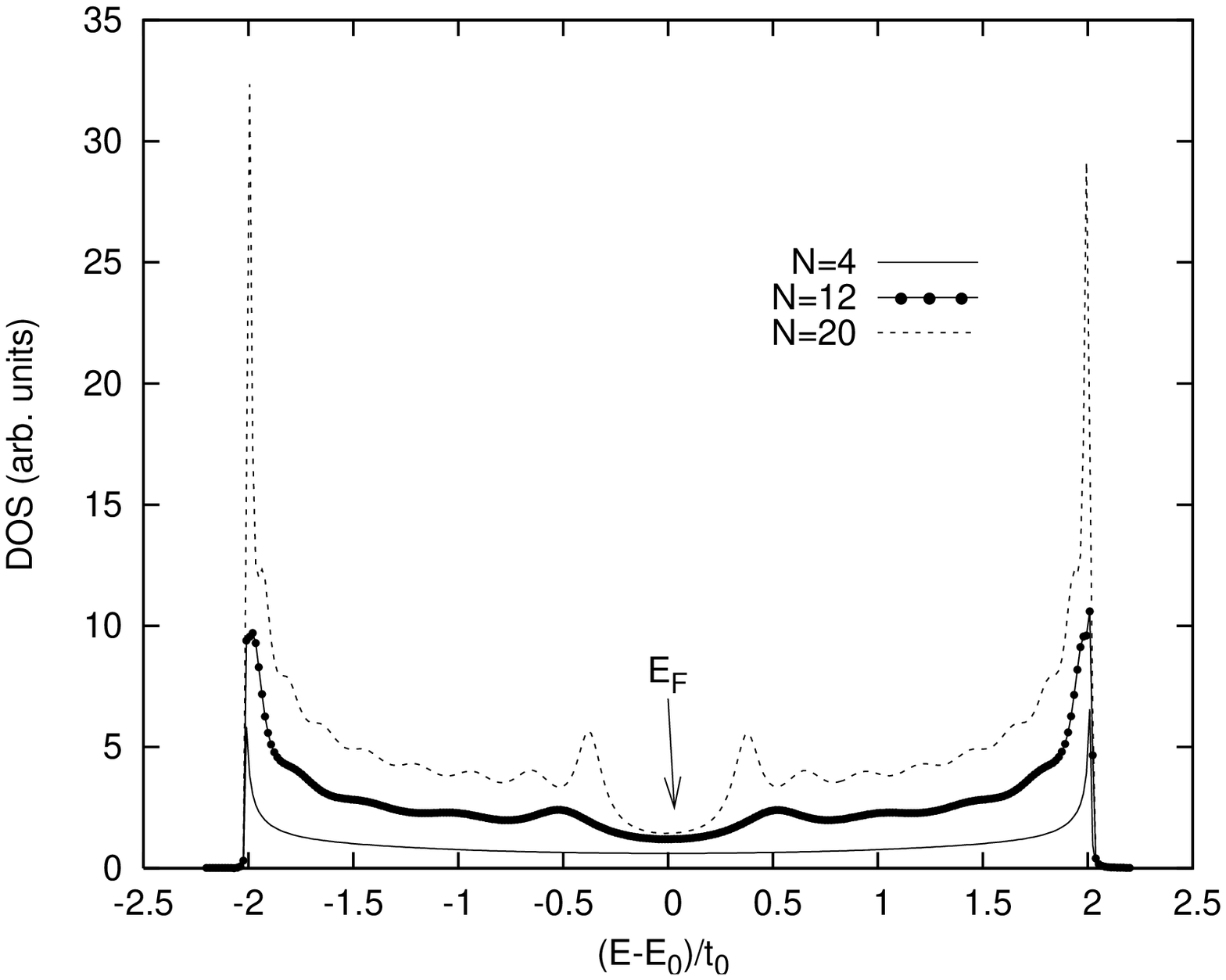}
\includegraphics[width=0.49\linewidth]{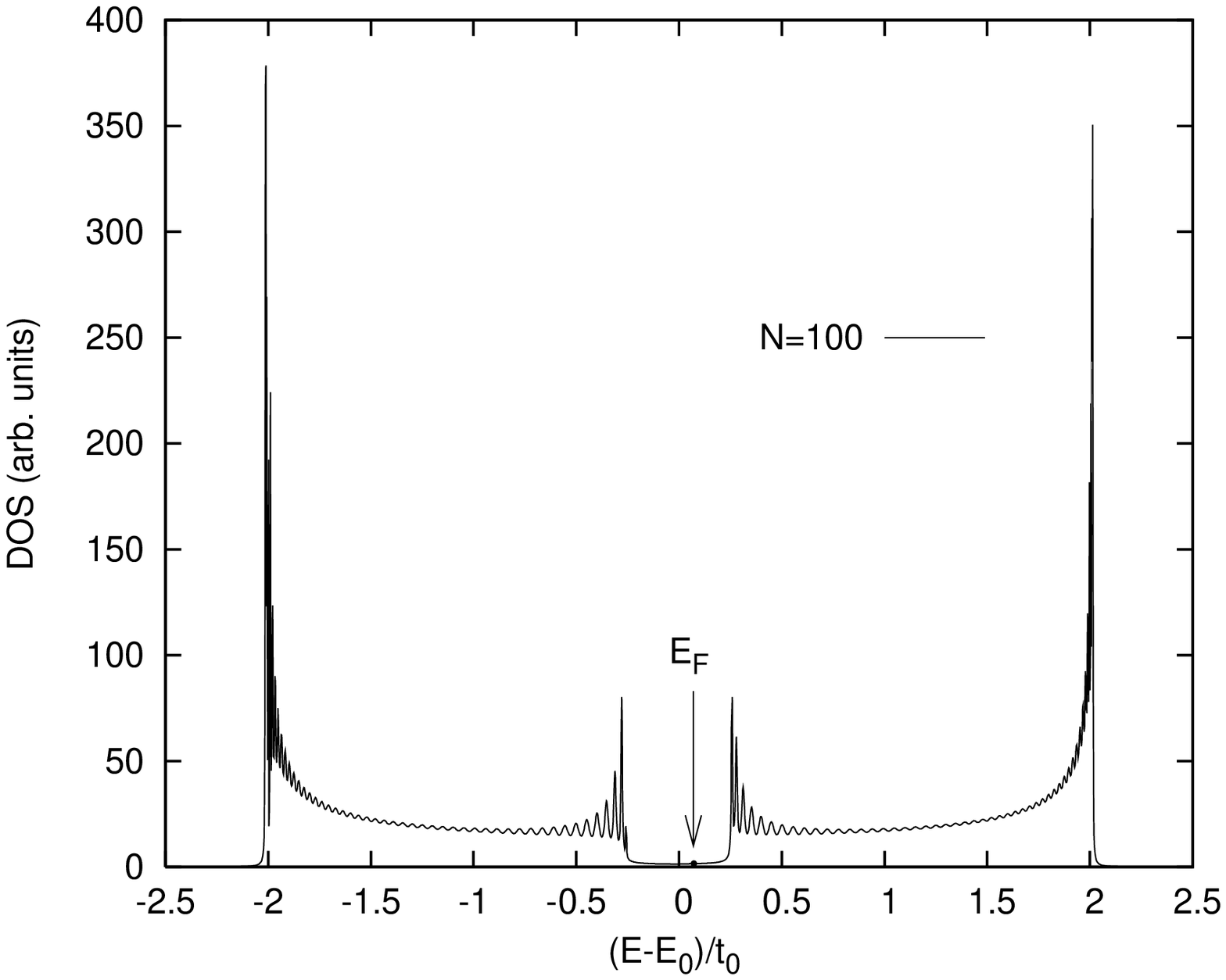}
\caption{DOS of periodic QDA connected to an incoming and 
an outgoing lead, as a function of energy
$E$ referred to the band center at energy $E_0$ and for a 
bandwidth of $4t_0$. Left panel: $N=4$ (solid line), $N=12$ (solid line and
dots) and 
$N=20$ (dashed line). Right panel: $N=100$.\label{fig:dos3}}
\end{center}
\end{figure}

\begin{figure}
\begin{center}
\includegraphics[width=0.49\linewidth]{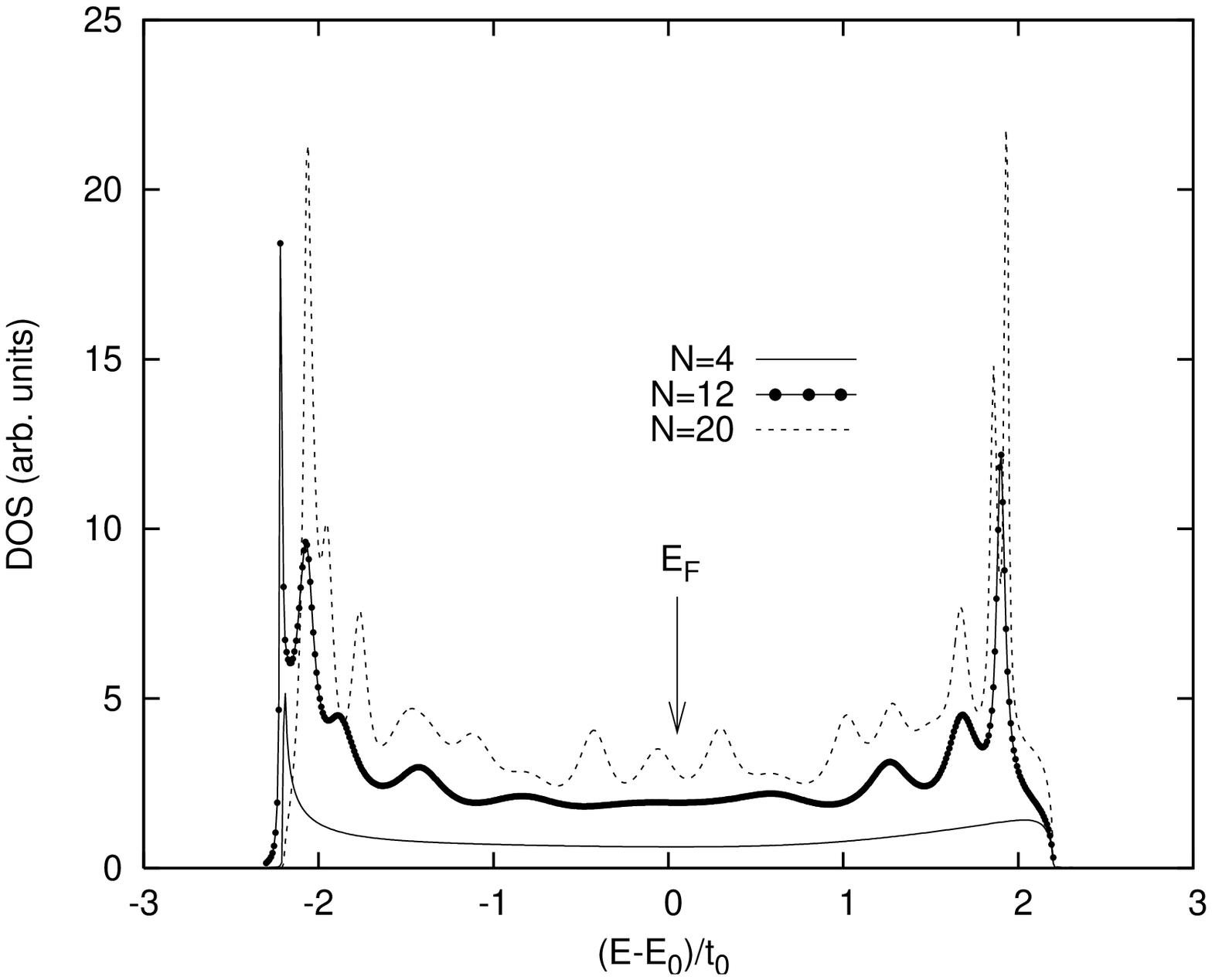}
\includegraphics[width=0.49\linewidth]{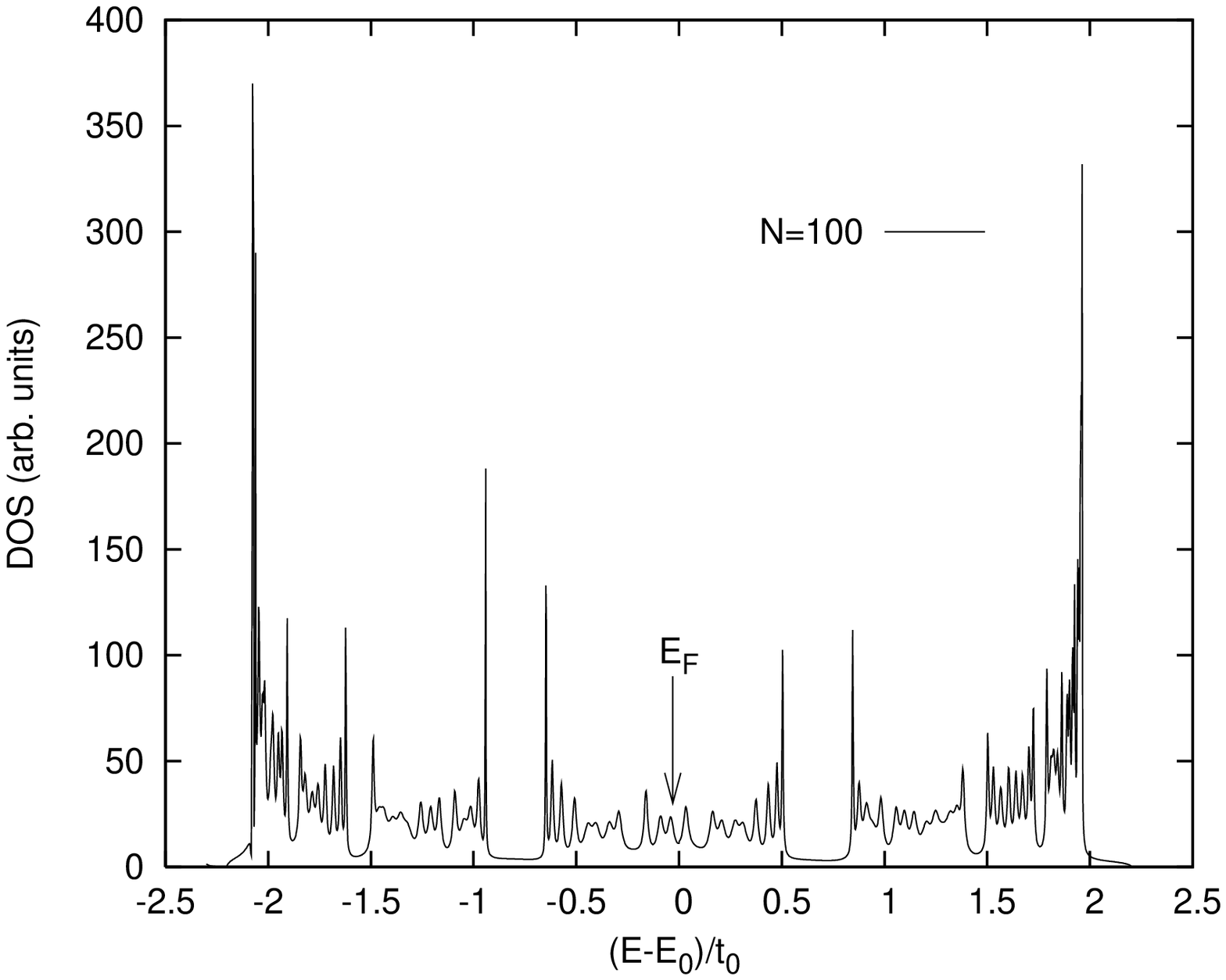}
\caption{DOS as in Fig. \ref{fig:dos3}, for a
Fibonacci-ordered QDA connected to an incoming and 
an outgoing lead.\label{fig:dos4}}
\end{center}
\end{figure}

The DOS of a periodic QDA connected to an incoming and an outgoing lead
is shown in Fig. \ref{fig:dos3}. For small numbers of QD's
($N=4$ and $N=12$) the effect of the leads is dominant
and the DOS resembles that of a 1D monatomic lattice.
A depression appears at the
center of the spectrum for $N=20$, and for a long array ($N=100$)
the effect of the leads becomes irrelevant and the DOS is very similar
to that shown in the left panel of Fig. \ref{fig:dos2} for the
isolated QDA. 
Similar effects on the DOS of the Fibonacci-ordered QDA are seen
 in Fig. \ref{fig:dos4}:
at low $N$ the peak structure shown in Fig. \ref{fig:dos1} is 
washed out by the presence of the leads, while at large $N$ the 
typical features due to quasi-periodicity appear.

At half filling the Fermi level is located near $(E-E_0)/t_0= 0$
for all configurations,
as is indicated in Figs. \ref{fig:dos3} and  \ref{fig:dos4}
by an arrow.
In this case the DOS 
at the Fermi level
vanishes on increasing the length of the periodic QDA. 
In a Fibonacci-ordered array, instead, 
the number of states at the Fermi level remains finite.
  
\section{Transmission through the array}
\label{trans_chain}
We rewrite the Hamiltonian
$H=\tilde H_{\textrm{QD}}+H_{\textrm{L,l}}+H_{\textrm{L,r}}$
of the effective dimer
representing the QDA connected to the leads
as the sum of two terms,
$H=H_0+H_I$.
The first term $H_0$ describes 
an infinite perfect chain with
spacing $a$, site energy $E_0$ and hopping energy $t_0$, whose elements
$(2,3,\dots,N-1)$ have been decimated as previously described,
and is given by
\begin{equation}
\hspace{-0.5cm}
H_0=H_{\textrm{L,l}}+H_{\textrm{L,r}}+(\tilde E_0-E_0)\,
\left(|1\rangle\langle1|+ (|N\rangle\langle N|)+\tilde t_0
(|1\rangle\langle N| +|N\rangle\langle 1|\right).
\end{equation}
The remainder
\begin{equation}
H_\textrm{I}=\tilde H_{\textrm{QD}}-\{\tilde E_0
(|1\rangle\langle1|+ (|N\rangle\langle N|)+\tilde t_0
(|1\rangle\langle N| +|N\rangle\langle 1|)\}
\end{equation}
will be viewed as a ``perturbation'' determining scattering
of incoming waves. In these equations
the quantities
$\tilde E_0$ and $\tilde t_0$ are obtained from Eqs. (\ref{eq:rec1})
and (\ref{eq:rec3}) for $j=N-2$ by taking 
$e_i=E_0$ and $t_{i,i+1}=t_0$
for all $i$.

The wavefunction $|\varphi\rangle$ at energy $E$ in the
continuous spectrum of $H$ is obtained from the wavefunction 
$|k\rangle$ of the unperturbed periodic Hamiltonian
$H_0$, the unperturbed Green's function $G^0(E)=(E-H_0)^{-1}$, and
the $T$ matrix
$T(E)=H_\textrm{I}(\textbf{1}-G^0H_\textrm{I})^{-1}$ as 
\begin{equation}
|\varphi\rangle=|k\rangle+G^0\,T\,|k\rangle\,,
\end{equation}
$T$ as well as $H_\textrm{I}$ being $2\times 2$ matrices in the space
spanned by $|0\rangle$ and $|N\rangle$.
The projection of $|\varphi\rangle$ onto the localized function
$|n\rangle$ is
\begin{equation}\label{eq:G}
\langle n|\varphi\rangle=\langle
n|k\rangle+\sum_{l,m=0,1}G^0_{n,l}\,T_{l,m}\,\langle m|k\rangle\,,
\end{equation}
where $G^0_{n,l}=\langle n|G^0|l\rangle$, $T_{l,m}=\langle
l|T|m\rangle$, and $\langle m | k \rangle=e^{i k m a}$. The
expressions for these matrix elements are given in \cite{patty}.
We can then write Eq. (\ref{eq:G}) in the form
\begin{equation}
\hspace{-0.6cm}
\langle n|\varphi\rangle=e^{i k n
a}+\left(G^0_{N,1}T_{1,N}+G^0_{1,N}T_{N,1}e^{-2ik(N-1)a}+G^0_{N,N}T_{N,N}+G^0_{1,1}T_{1,1}\right)e^{i
k n a}
\end{equation}
and define the transmittance $\tau$ and the reflectance
$\rho$ by writing
\begin{displaymath}
\langle n|\varphi\rangle = \left\{
 \begin{array}{ll}
\tau e^{ikna} & (n>1)\\
e^{ikna}+\rho\, e^{-ikna} & (n<0)
\end{array} \right.\,.
\end{displaymath}
We thus obtain
\begin{equation}
\tau=1+G^0_{N,1}T_{1,N}+G^0_{1,N}T_{N,1}e^{-2ik(N-1)a}+G^0_{N,N}T_{N,N}+G^0_{1,1}T_{1,1}.
\label{eq_diff}
\end{equation}
Notice the difference between this Eq. (\ref{eq_diff}), which is valid for
both real and effective dimers, and Eq. (8) in Ref. \cite{patty}
applies only to real dimers.

\begin{figure}
\begin{center}
\includegraphics[width=0.49\linewidth]{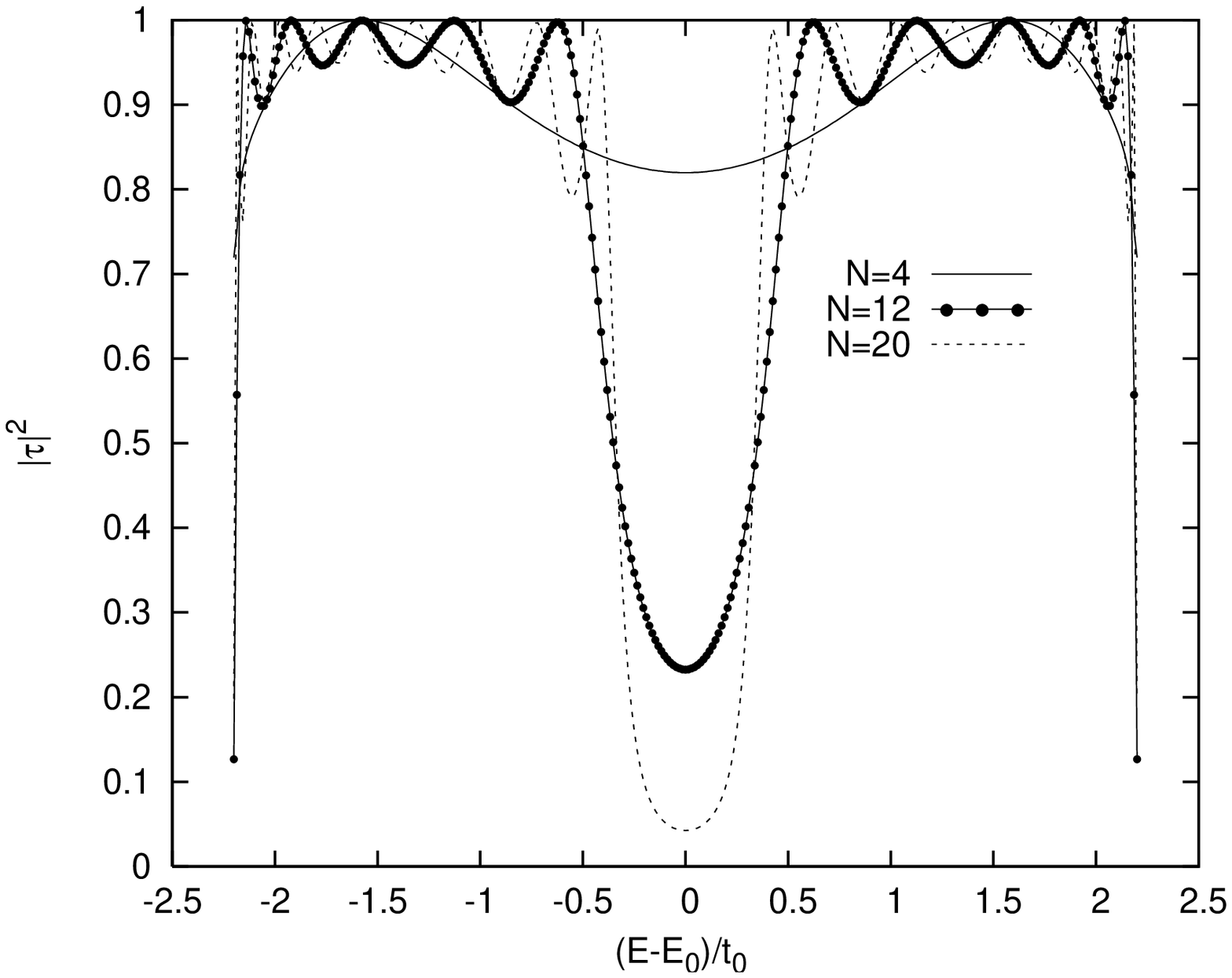}
\includegraphics[width=0.49\linewidth]{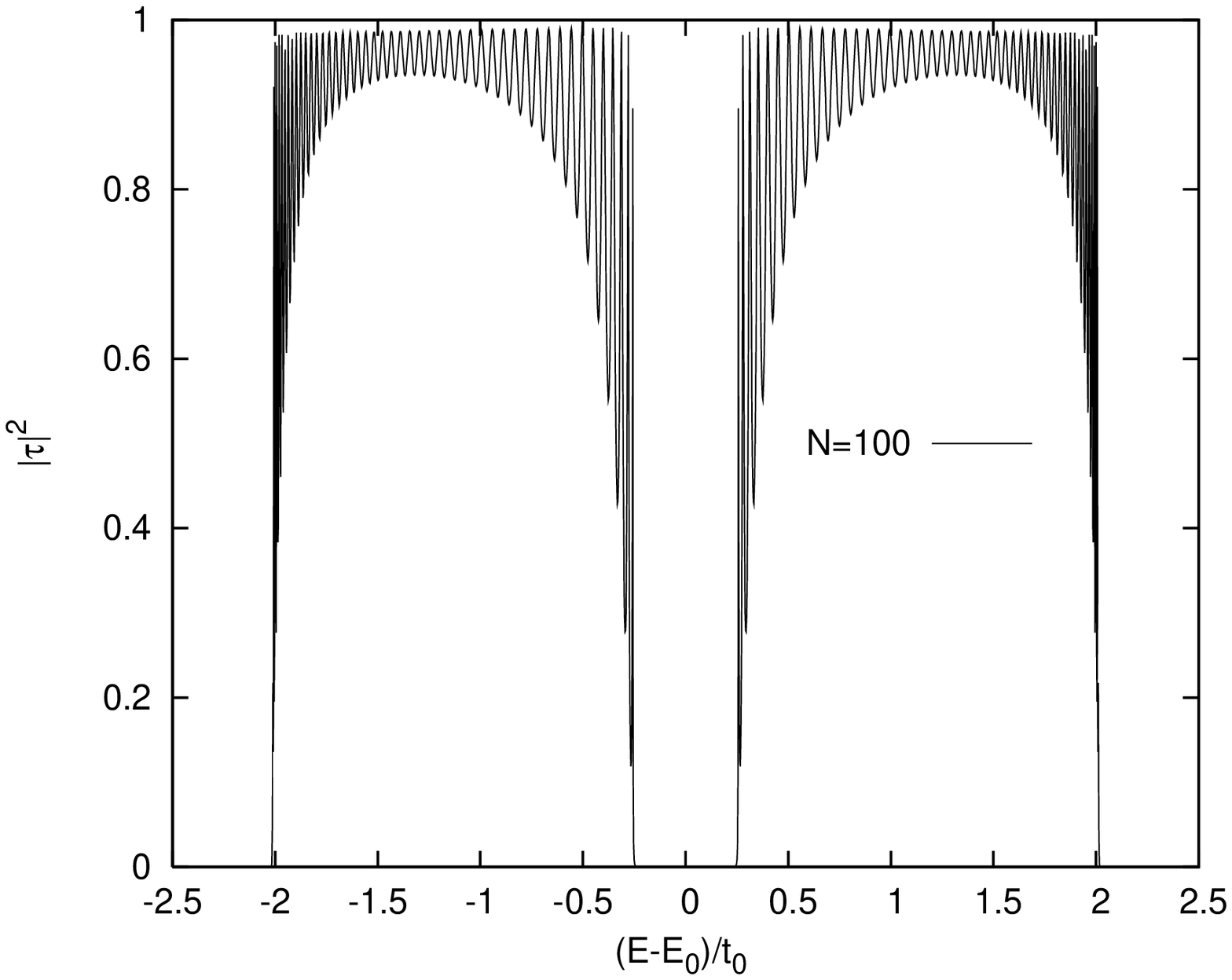}
\end{center}
\caption{Transmittivity coefficient for a periodic QDA
as a function of the energy $E$,
referred to the band center at energy $E_0$ and for a 
bandwidth of $4t_0$. 
Left panel: $N=4$ (solid line), $N=12$ (solid line and
dots) and $N=20$ (dashed line). Right panel: $N=100$.}
\label{fig:tr}
\end{figure}

\begin{figure}
\begin{center}
\includegraphics[width=0.49\linewidth]{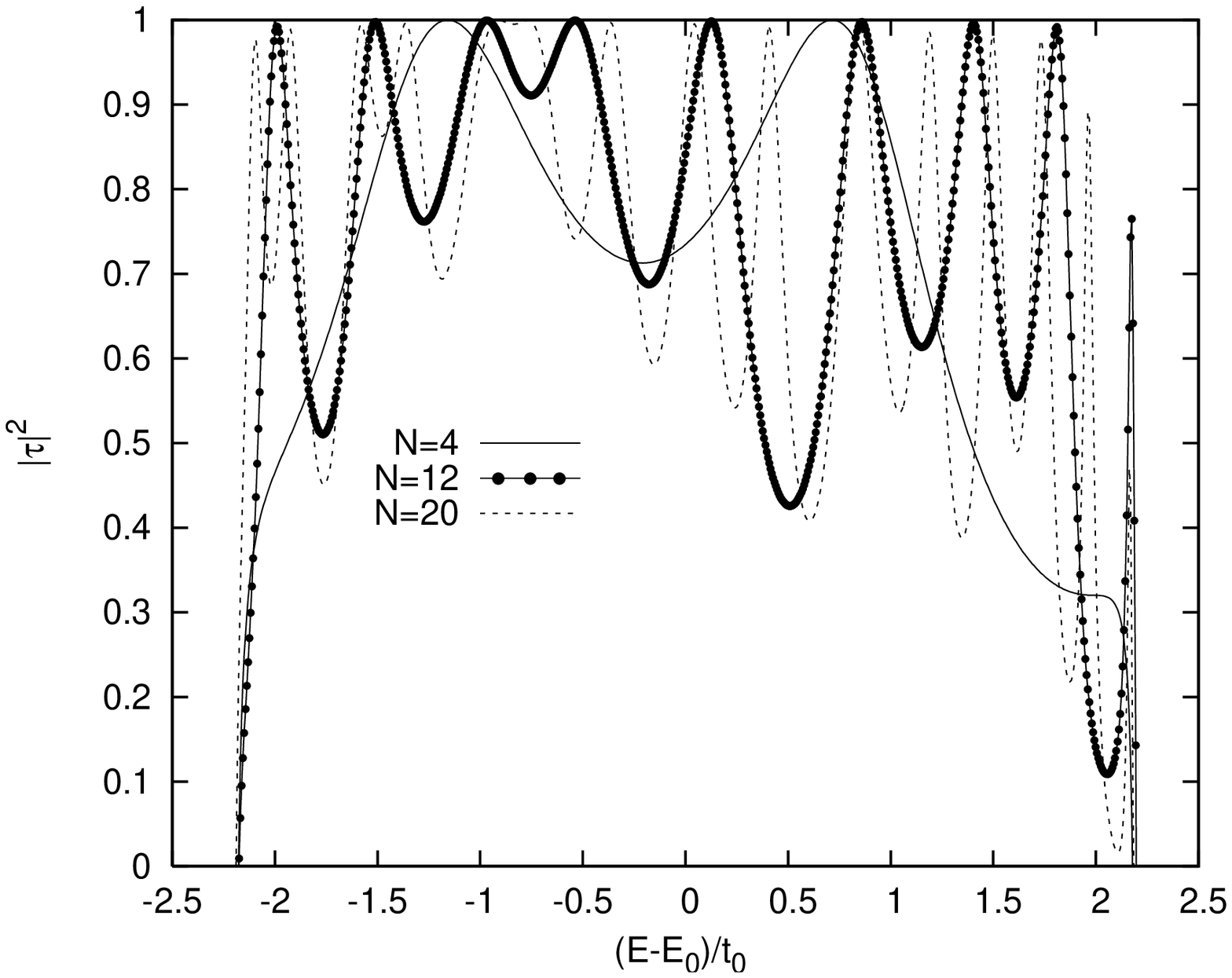}
\includegraphics[width=0.49\linewidth]{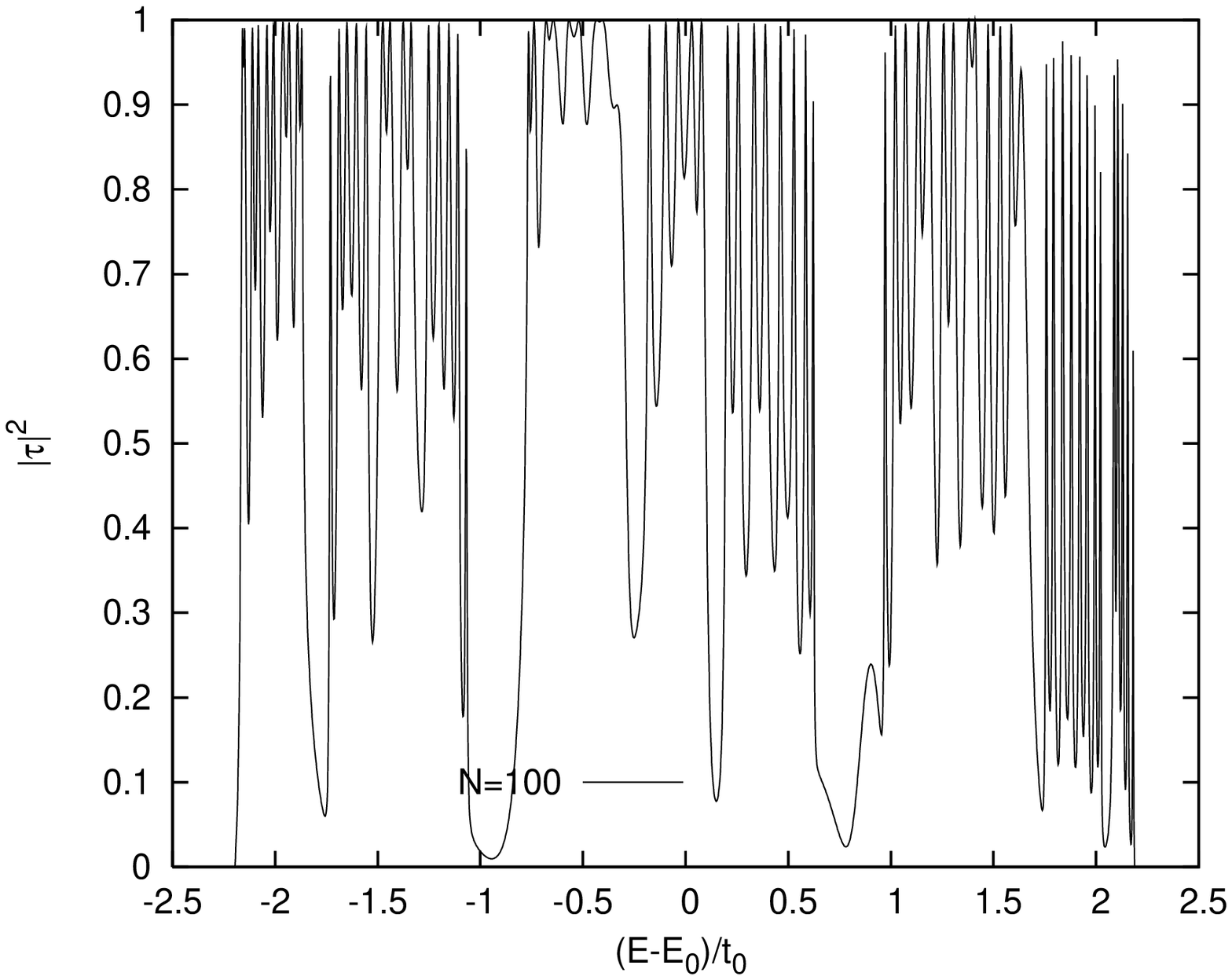}
\end{center}
\caption{Transmittivity coefficient as in Fig. \ref{fig:tr},
for a Fibonacci-ordered array.}
\label{fig:tr1}
\end{figure}

The transmission coefficient $\mathcal{T}$ is given by
$\mathcal{T}=|\tau|^2$.
Numerical results for the periodic and the Fibonacci-ordered
QDA are shown in Figs. \ref{fig:tr} and \ref{fig:tr1}. 
The features of the transmittivity as a function of the energy 
parallel those of the
DOS in Figs. \ref{fig:dos3} and \ref{fig:dos4}, where the effects
of the leads have been included:
the transmittivity drops as the DOS decreases and
vanishes in the energy gaps.
At half filling in the periodic array
the transmittivity
at the Fermi level is high for very short arrays but rapidly drops 
on increasing the number of dots, so that the array
becomes insulating.
On the contrary, in a Fibonacci-ordered array the transmittivity 
near the Fermi level remains high, so that metallic-like conduction 
is preserved independently of the length
of the array. There are, however, strong minima in the transmittivity in 
correspondence to the pseudo-gaps in the DOS.

\section{Concluding remarks}
\label{concl}
In summary,
we have presented a fully analytical method for the
calculation of the density of electron states and the 
phase-coherent electron transport coefficient in a linear array
made of an arbitrary number of two different types of potential well
with different energy levels. We have specifically considered
both a periodic sequence of pairs of potential wells and a
sequence in which the two types of potential wells are arranged
according to the Fibonacci sequence. In both cases the
connection of the array to external metallic wires has been included.
The method does not
explicitly diagonalize  the tight-binding Hamiltonian of the system, 
but uses a decimation-renormalization procedure to reduce
the array to an effective dimer, which is then treated by a suitably
adapted $T$-matrix
scattering approach.

The density of states and the transmittivity vanish at 
the same energy when the connection to external metallic wires is
included. The calculations explicitly show that an energy gap
opens in the periodic array as its length is increased, so that
metallic conductance turns into insulating behavior in the case 
of single occupancy of each well. With increasing length the 
Fibonacci-ordered array develops instead pseudo-gaps away from the 
band center, and again a metal-to-insulator transition is
expected for appropriate values of the filling factor.

\ack
Partial support by MIUR under the PRIN2003 Program is gratefully acknowledged.
One of us (M.R.B.) thanks Dr. Reza Asgari for fruitful discussions.


\begin{thebibliography}{99}
\bibitem{datta}
S. Datta, Electronic Transport in Mesoscopic Systems, 
Cambridge University Press, Cambridge, 1995. 
\bibitem{ahmed}
H. Ahmed, M. Pepper, A. Broers, Transport in Nanostructure,
Cambridge University Press, Cambridge, 1997. 
\bibitem{landauer}
R. Landauer, Philos. Mag. 21 (1970) 863.
\bibitem{jacak}
L. Jacak, P. Hawrylak, A. W\`{o}js, Quantum Dots, Springer, 
Berlin, 1998.
\bibitem{ref2}T. Chakraborty, Quantum Dots: A Survey of the Properties 
of Artificial Atoms, Elsevier, Amsterdam, 1999.
\bibitem{tarucha}
S. Tarucha, D.G. Austing, T. Honda, R.J. van der Hage, L.P. Kouwenhoven,
Phys. Rev. Lett. 77 (1996) 3613.
\bibitem{Burkard}
G. Burkard, H. Engel, D. Loss,
Fortschritte Phys. 48  (2000) 965.
\bibitem{tews}
M. Tews, Ann. Phys. (Leipzig) 12 (2004) 249.
\bibitem{aldea} 
V. Moldoveanu, A. Aldea, A. Manolescu, M. Nita,
Phys. Rev. B 63 (2001) 045301. 
%Magneoconductance in chaotic quantum dots 2D arrays
\bibitem{louis} E. Louis, J. A. Verges,
Phys. Rev. B 63 (2001) 115310.
%Kondo resonances and Fano antiresonances in transport through quantum dots
\bibitem{torio}
 M. E. Torio, K. Hallberg, A. H. Ceccatto, C. R. Proetto, 
Phys. Rev. B 65 (2002) 085302.
%Electronic transport through a quantum dot network
\bibitem{dorn} A. Dorn, T. Ihn, K. Ensslin, 
W. Wegscheider, M. Bichler, Phys. Rev. B. 70 (2004) 205306.
\bibitem{Kirczenow}
G. Kirczenow, Phys. Rev. B 46 (1992) 1439.
\bibitem{tanatar}
V. Moldoveanu, A. Aldea, B. Tanatar,
Phys. Rev. B 70 (2004) 085303. 
\bibitem{mardani}
M. Mardaani, K. Esfarjani, {\tt cond-mat}/0403190.
\bibitem{mardani2}
M. Mardaani, K. Esfarjani, Physica E 15 (2004) 119.
%\bibitem{tosi}
%Y. Eksioglu, P. Vignolo and M.P. Tosi,\textit{ cond-mat}/0404458
%\bibitem{schet}
%D. Schetman, I. Blech, D. Gratias, and J.W Cahn. Phys. Rev. Lett.
%{\bf 53}, 1951 (1984).
\bibitem{grosso}
R. Farchioni, G. Grosso, G.P. Parravicini. Phys. Rev. B. 45
(1992) 6368.
\bibitem{patty}
R. Farchioni, P. Vignolo, G. Grosso. Phys. Rev. B. 60 (1999) 15705.
\bibitem{kirkman}
P.D. Kirkman, J.B. Pendry, J. Phys. C 17 (1984) 4327.
\bibitem{patty_k}
R. Farchioni, G. Grosso, P. Vignolo
Phys. Rev. B 62 (2000) 12565.
\bibitem{zehra}
P. Vignolo, Z. Akdeniz, M.P. Tosi, J. Phys. B 36 (2003) 4535.
\end{thebibliography}
\end{document}